\documentclass[conference]{IEEEtran}
\IEEEoverridecommandlockouts
\usepackage{cite}
\usepackage{amsmath,amssymb,amsfonts}
\usepackage{algorithmic}
\usepackage{graphicx}
\usepackage{textcomp}
\usepackage{xcolor}
\usepackage{braket}
\usepackage{hyperref}
\usepackage{tabularx}
\usepackage{threeparttable}
\usepackage{multirow}
\usepackage{booktabs} 
\usepackage{subcaption}
\usepackage{tikz}
\usetikzlibrary{shapes.geometric, arrows.meta, positioning, calc, shadows}

\definecolor{boxblue}{RGB}{235, 245, 255}
\definecolor{borderblue}{RGB}{0, 102, 204}
\definecolor{textgray}{RGB}{50, 50, 50}

\def\BibTeX{{\rm B\kern-.05em{\sc i\kern-.025em b}\kern-.08em
    T\kern-.1667em\lower.7ex\hbox{E}\kern-.125emX}}

\hypersetup{ 
    pdfauthor={Zhehao Yi, Rahul Bhadani},   
    pdftitle={Q-LINK: Quantum Layerwise Information Residual Network via a Messenger Qubit for Barren Plateaus Mitigation},   
    pdfsubject={Quantum Machine Learning}, 
    pdfkeywords={Variational Quantum Circuits},
    pdfcreator={Zhehao Yi}, 
    pdfproducer={Zhehao Yi}, 
     colorlinks=true,
     filecolor=green, 
     citecolor = blue,       
     urlcolor=red, 
} 

\begin{document}

\title{Q-LINK: Quantum Layerwise Information Residual Network via a Messenger Qubit for Barren Plateaus Mitigation\\
}

\author{
\IEEEauthorblockN{1\textsuperscript{st} Zhehao Yi}
\IEEEauthorblockA{\textit{AI, Autonomy, Resilience, Control (AARC) Lab}\\\textit{Electrical \& Computer Engineering} \\
\textit{The University of Alabama in Huntsville}\\
Huntsville, United States \\
zhehao.yi@uah.edu}
\and
\IEEEauthorblockN{2\textsuperscript{nd} Rahul Bhadani}
\IEEEauthorblockA{\textit{AI, Autonomy, Resilience, Control (AARC) Lab}\\\textit{Electrical \& Computer Engineering} \\
\textit{The University of Alabama in Huntsville}\\
Huntsville, United States \\
rahul.bhadani@uah.edu}
\and
}

\maketitle

\begin{abstract}
In hybrid classical–quantum computing, variational quantum algorithms (VQAs) have emerged as a promising approach in the Noisy Intermediate-Scale Quantum (NISQ) era; however, their performance is often hindered by barren plateaus, where gradients vanish exponentially, rendering optimization ineffective. In this work, we introduce a residual-inspired quantum circuit architecture that incorporates a single messenger qubit, referred to as Q-LINK. By conducting numerical simulations on random quantum states, we observe that Q-LINK significantly enhances optimization behavior by sustaining larger gradient variance and accelerating convergence. Additionally, Q-LINK improves convergence efficiency by 4–6 times and increases gradient variance by up to two orders of magnitude compared with the Vanilla model. To further characterize the impact of the proposed structure, we analyze the expressibility of the circuits before and after introducing Q-LINK and find that the overall expressibility value remains largely unchanged, indicating that the original representational capacity of the circuit is preserved. In addition, we visualize the loss landscapes of different architectures to provide insights into how the proposed design reshapes the cost function landscape. These results demonstrate that introducing only a single messenger qubit can effectively mitigate barren plateau effects while maintaining the ability to explore the Hilbert space of variational quantum circuits.
\end{abstract}


\section{Introduction}
Quantum computing is an emerging technology bridging the gap between quantum mechanics and practical computing applications. It has shown strong potential in handling high-dimensional data and has been applied in various domains, such as quantum chemistry \cite{mustafa2022}, quantum optimization \cite{sciorilli2025}, and quantum finance \cite{innan2025}. Most of the existing quantum machine learning algorithms adopt a hybrid quantum-classical construction, enabling a minimal quantum resource to be made useful with classical optimization \cite{mcclean2016}.

In quantum machine learning, variational quantum algorithms (VQAs) serve as a core framework \cite{cerezo2021}. A VQA consists of a quantum part and a classical part. The quantum part includes a quantum encoding process and a parameterized quantum circuit, where the quantum circuit serves as the core implementation of the quantum algorithm; the classical part consists of an optimizer that updates these parameters by minimizing a cost function, typically using gradient descent optimization methods. This hybrid quantum–classical structure is similar to classical neural networks and is well-suited for the Noisy Intermediate-Scale Quantum (NISQ) era. As a result, VQAs have been widely explored in many applications, including protein structure optimization \cite{robert2021} and path planning \cite{sarkar2024}.

Despite their advantages, the trainability of VQAs is severely affected by the barren plateau problem \cite{larocca2025}. This problem occurs when the number of qubits or circuit depth increases, and gradients can decay exponentially, leading to slow or stalled training. To mitigate this problem, several methods have been proposed, such as neural network-based parameter or state generation \cite{friedrich2022, yi2025nn, yi2024}, entanglement-aware learning strategies \cite{patti2021}, state-efficient ansatz \cite{liu2024}, and classical controllers for parameter updates \cite{yi2026}.

Recently, a residual-based strategy has been introduced to mitigate barren plateaus by dividing the circuit into blocks and combining measurement outcomes to form a residual structure \cite{kashif2024}. However, intermediate measurements may disrupt quantum coherence and entanglement due to measurement-induced state collapse \cite{nielsen2010}. Motivated by this limitation, we propose a Quantum Layerwise Information Residual Network via a Messenger Qubit (Q-LINK) method. By introducing a messenger qubit to collect and redistribute information from data qubits without intermediate measurements, the residual-like information pathway preserves the unitary structure of the quantum circuit and maintains the quantum coherence and entanglement.

In this paper, we introduce two Q-LINK models. When the collection parameter is fixed to $\pi/4$, the model is referred to as Q-LINK (Fixed), while the model with trainable parameters optimized via a cost function is denoted as Q-LINK (Adaptive), and the model without the messenger qubit is referred to as the Vanilla model. We compare the convergence behavior of Q-LINK and Vanilla models to demonstrate the effectiveness of the messenger qubit. Meanwhile, we visualize the loss landscape of different models, offering a more intuitive way to understand the optimization. Finally, we calculate the expressibility and gradient variance of the three models to further analyze their training dynamics.

This article is organized as follows: section \ref{sec2} provides the necessary background information, section \ref{sec3} describes the proposed methodology, section \ref{sec4} illustrates and analyzes the simulation results, and finally section \ref{sec5} summarizes the key findings of this study and future work.

\section{Background}
\label{sec2}
\subsection{Variational Quantum Algorithm}
\begin{figure}[htbp]
    \centering
    \includegraphics[width=0.9\linewidth]{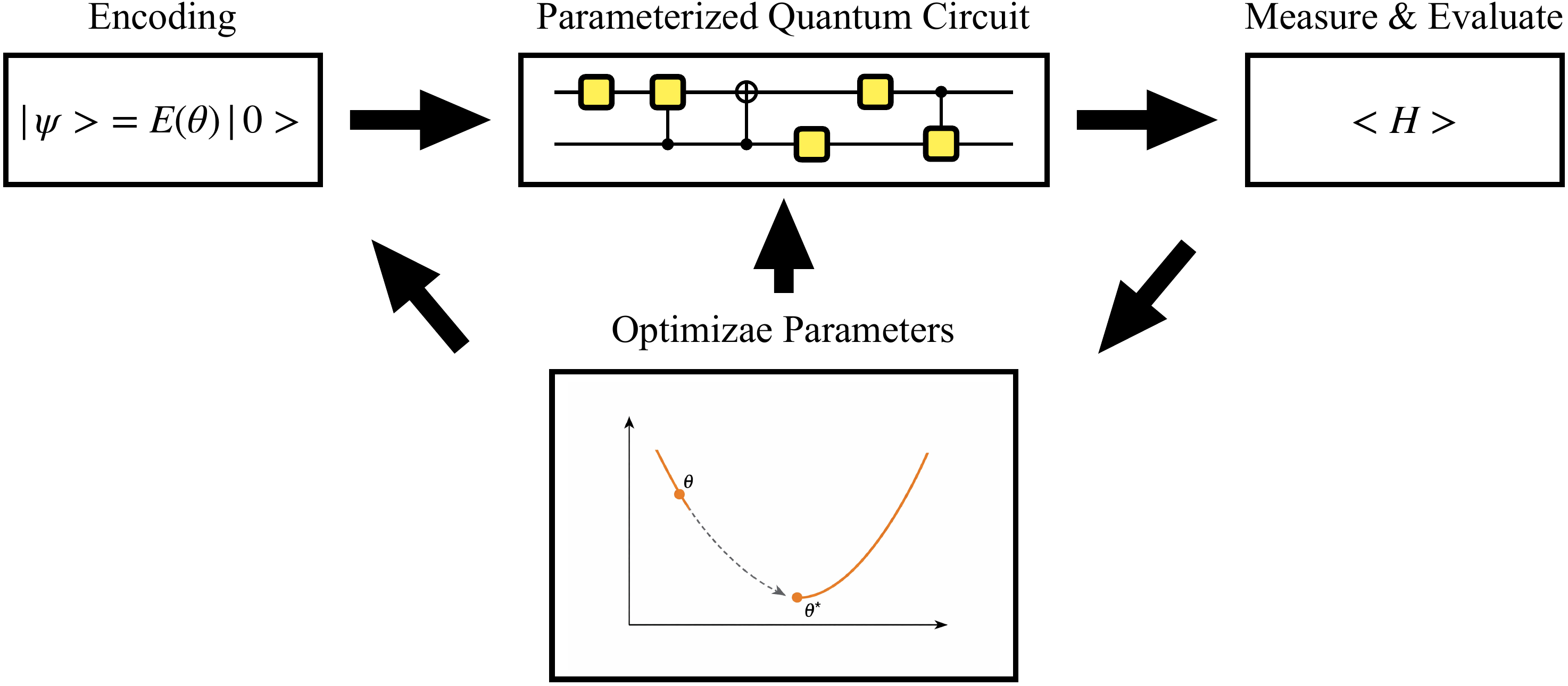}
    \caption{Workflow of a variational quantum algorithm. An initial ground state is encoded into a quantum state, followed by iterative execution of a parameterized quantum circuit. Measurement results are used to compute the cost function, and a classical gradient-based optimizer updates the parameters of the encoding and circuit until convergence.}
    \label{fig: vqa}
\end{figure}

 A variational quantum algorithm consists of four main components: encoding, circuit operation, measurement, cost function evaluation, and parameter update. First, the ground state $\ket{0}$ is transformed into an input quantum state $\ket{\psi}$ through a specific encoding process. The operation of the quantum circuit can then be viewed as a sequence of unitary multiplication, which is collectively represented by a unitary operator $U(\theta)$, where $\theta$ denotes a sequence of trainable parameters. Next, measurements are performed on the output quantum state, and the resulting expectation values are used to construct a predefined cost function $\mathcal{L}$. The computed loss value is then employed by a classical optimizer to update the circuit parameters; the overall procedure is illustrated in Figure \ref{fig: vqa}.

\subsection{Trainability of VQA}
Since variational quantum circuits typically rely on gradient-based optimization to update their parameters, computing parameter gradients with respect to the cost function $\frac{\partial\mathcal{L}}{\partial \theta}$ can lead to an exponential suppression of the gradient magnitude $\frac{\partial\mathcal{L}}{\partial \theta} \leq O(\frac{1}{a^n}), a \geq 1$, causing it to rapidly approach zero as the system size increases \cite{friedrich2022}. This phenomenon is commonly referred to as \textbf{the barren plateau problem}. 

To quantify the maximum extent to which a variational quantum circuit can explore the Hilbert space, the concept of circuit expressibility was introduced, measuring the similarity between the distribution of states generated by the quantum circuit and the Haar-random state distribution \cite{sim2019}. Expressibility is evaluated using the Kullback–Leibler (KL) divergence between the probability distribution generated by the circuit states $P_{pqc}$ and that of the Haar-random quantum state $P_{haar}$, shown in Equation \eqref{eq: expr}. 
\begin{equation}
    \mathrm{Expresibility} = D_{\mathrm{KL}}(P_{pqc} || P_{haar})
    \label{eq: expr}
\end{equation}
To obtain the circuit state probability distribution, the quantum circuit is fixed after training, and a histogram-based method is employed to approximate the distribution. The probability distribution of Haar-random states is known analytically and can be expressed as $P_{haar} = (N-1)(1-F)^{N-2}$, where $F$ denotes the fidelity between two quantum states and $N$ is the dimension of the Hilbert space. A smaller KL divergence indicates that the circuit more closely approximates a random vector in the Hilbert space, corresponding to stronger expressibility.

\subsection{Residual Structure}
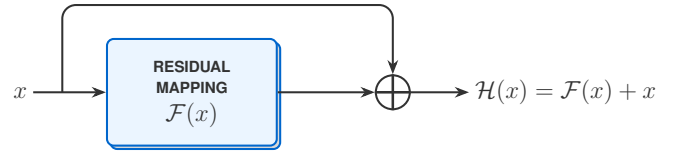
\begin{figure}[htbp]
    \centering
\resizebox{\linewidth}{!}{
\begin{tikzpicture}[
        node distance=2.5cm,
        font=\sffamily,
        >=Stealth, 
        block/.style={
            rectangle, 
            draw=borderblue, 
            fill=boxblue, 
            very thick, 
            rounded corners=4pt,
            minimum width=3.5cm, 
            minimum height=2.2cm, 
            text centered, 
            copy shadow={shadow xshift=2pt, shadow yshift=-2pt, fill=black!15},
            text=textgray
        },
        plus/.style={
            circle, 
            draw=textgray, 
            fill=white,
            very thick, 
            minimum size=0.7cm, 
            inner sep=0pt
        },
        line/.style={draw=textgray, line width=1.2pt, ->}
    ]

        \node (input) [text=textgray] {\Large $x$};
        
        \node [block, right=1.5cm of input] (residual) {
            \begin{tabular}{c} 
                \textbf{\small RESIDUAL} \\ \textbf{\small MAPPING} \\ \addlinespace[2pt] \Large $\mathcal{F}(x)$ 
            \end{tabular}
        };
        
        \node [plus, right=2cm of residual] (add) {};
        \draw [line width=1.5pt, textgray] (add.north) -- (add.south);
        \draw [line width=1.5pt, textgray] (add.west) -- (add.east);
        
        \node [right=1.2cm of add, text=textgray] (output) {\Large $\mathcal{H}(x) = \mathcal{F}(x) + x$};

        
        \draw [line] (input) -- (residual);
        \draw [line] (residual) -- (add);
        \draw [line] (add) -- (output);

        \coordinate (split) at ($(input.east)!0.4!(residual.west)$);
        \draw [line, rounded corners=8pt] (split) -- ++(0, 1.8) -| (add.north);

    \end{tikzpicture}
}
    \caption{Residual block structure in a classical neural network. The input is combined with the output of a nonlinear transformation through a skip connection, enabling direct information propagation across layers.}
    \label{fig: residual}
\end{figure}

In classical machine learning, deep neural networks are often more difficult to train due to information loss or ineffective information propagation during training. To address this issue, residual learning was proposed \cite{he2016}. In a residual neural network, assuming that the input and output dimensions are the same, the network does not directly approximate the mapping $\mathcal{H}(x)$. Instead, the nonlinear layers are designed to approximate a residual mapping $\mathcal{F}(x) = \mathcal{H}(x)-x$, such that the original mapping can be expressed as $\mathcal{H}(x) = \mathcal{F}(x) + x$. The module that implements this transformation is denoted as a residual block, as illustrated in Figure \ref{fig: residual}.
\section{Method}
\label{sec3}
In this work, we use randomly initialized quantum states as inputs (data qubits) and construct variational quantum circuits with sufficient depth to exhibit a barren plateau. The circuit consists of multiple parameterized layers acting on data qubits.

We define the circuit without the messenger qubit as the Vanilla model throughout this work. This circuit serves as the baseline architecture for comparison with the proposed Q-LINK model. The Vanilla circuit only consists of operation layers. Each operation layer applies a sequence of single-qubit rotation gates, followed by entangling operations between qubits. Specifically, for each qubit, the rotation gates $R_z$, $R_y$, and $R_x$ are applied, after which controlled-Z ($CZ$) gates are applied between neighboring qubits. The circuit can be written as 
\begin{equation}
\resizebox{0.9\linewidth}{!}{$
U_{\mathrm{Vanilla}} =
\prod_{d=1}^{D}
\left(
\prod_{(i,j)\in \mathcal{E}} CZ_{i,j}
\right)
\left(
\bigotimes_{i=1}^{N}
R_x^{(i)}(\theta_{x}^{d,i})
R_y^{(i)}(\theta_{y}^{d,i})
R_z^{(i)}(\theta_{z}^{d,i})
\right)
$}
\label{eq:Vanilla}
\end{equation}
where $D$ and $N$ represent the circuit depth and the number of qubits, respectively.

\begin{figure*}[htbp]
    \centering
    \includegraphics[width=0.7\linewidth]{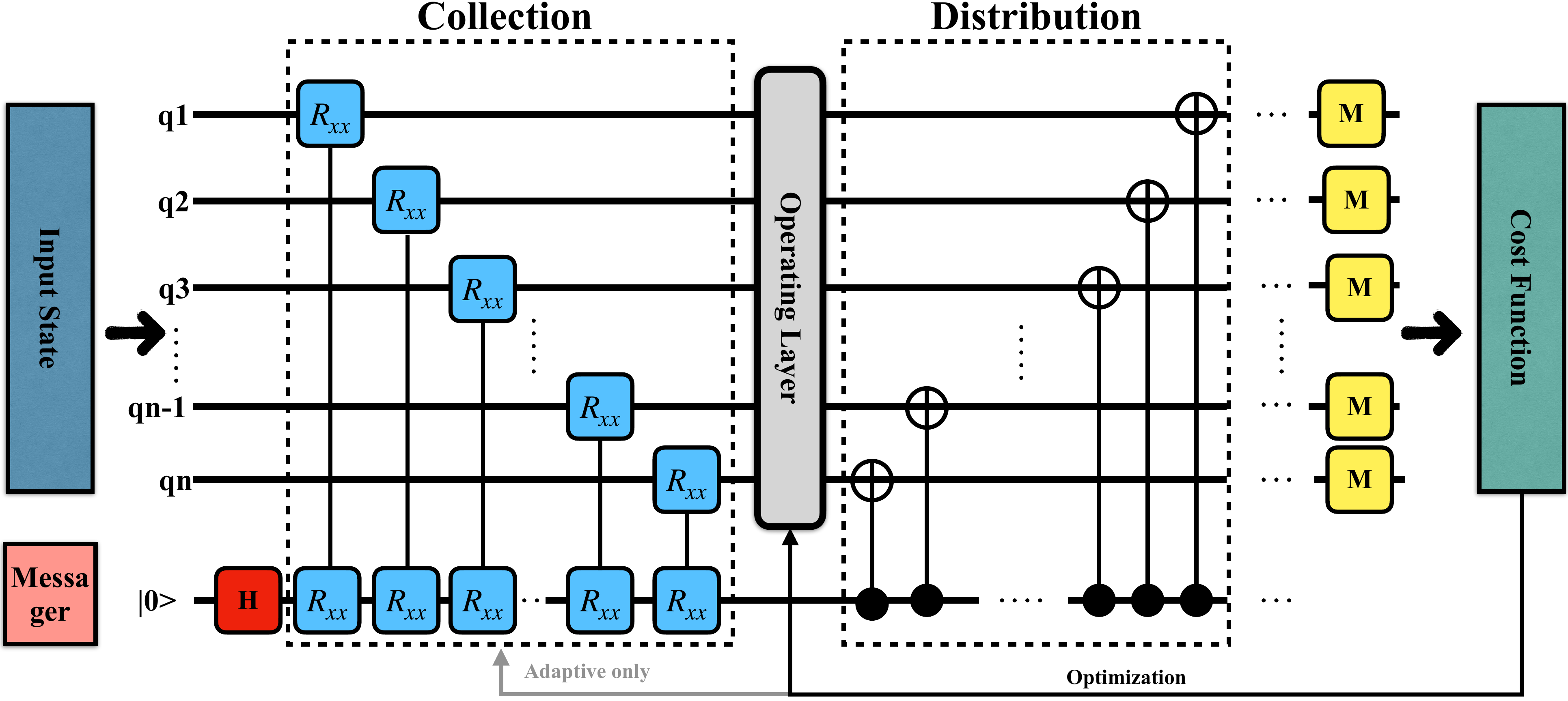}
    \caption{Detailed model structure of the Q-LINK. The circuit consists of a collection part and a distribution part mediated by a messenger qubit. Q-LINK includes a fixed variant with the $R_{xx}$ gate parameter set to $\pi/4$, denoted as Q-LINK (Fixed), and an adaptive variant with trainable $R_{xx}$ gate parameters, denoted as Q-LINK (Adaptive).}
    \label{fig: qlink}
\end{figure*}

To introduce a residual-like structure without using intermediate measurements, we propose the Q-LINK method, shown in Figure \ref{fig: qlink}. Q-LINK adds an extra messenger qubit to the circuit, which is initialized in the ground state $\ket{0}$.
The messenger qubit interacts with the data qubits during the circuit evolution and serves as a carrier of information, enabling layerwise residual information propagation.

The messenger residual mechanism contains two parts: information collection and information distribution. In the collection part, the messenger qubit interacts with each data qubit by using $R_{xx}$ gates, creating correlations between the data qubits and the messenger qubit before the operation layer. In the distribution part, after the data qubits undergo further operation layers, the messenger qubit interacts with the data qubits through CNOT gates, allowing information to be merged back into the data qubits. This process is repeated across circuit layers; it can be represented as $U_{\mathrm{Q-LINK}} = U_{\mathrm{dist}}U_{\mathrm{opr}}U_{\mathrm{coll}}$. This mechanism enables information captured before the operation layer to be reintroduced afterward, forming an identity-like information pathway similar to residual connections in classical neural networks.

We consider two different Q-LINK architectures. In Q-LINK (Fixed), the parameter of the $R_{xx}$ gates between the messenger qubit and the data qubits is fixed to $\pi/4$, and only the parameters in the data qubits layers are trainable. In contrast, Q-LINK (Adaptive) allows the $R_{xx}$ gate parameters to be trainable and optimized jointly with the data qubits parameters through the same cost function. At the end of the circuit, only the data qubits are measured to compute the cost function.

We adopt the same cost function used in previous studies \cite{friedrich2022} for the ground-state preparation task. In this work, we focus on the barren plateau problem in VQA. This ground-state preparation task is used as a benchmark, providing a controlled setting for evaluating circuit trainability.
\begin{equation}
\mathcal{L} = 1 - \frac{1}{n} \sum_{i=1}^{n} \braket{Z_i}
\label{eq:costfunction}
\end{equation}
where $\braket{Z_i}$ represents the expectation value of the $Z$ operator on the $i$-th qubit. Since the basis state $\ket{0}$ is the eigenstate of $Z$ when the eigenvalue is 1, maximizing $\braket{Z_i}$ encourages each qubit to converge toward $\ket{0}$. Therefore, minimizing $\mathcal{L}$ drives the circuit output toward the ground state $\ket{0}^{\otimes n}$.

Finally, we visualize the loss landscapes of different models \cite{li2018}. The parameters obtained after training are fixed and denoted as $\theta_{\mathrm{trained}}$. Then we selected two random directions $d_1$ and $d_2$, each having the same dimensionality as $\theta_{\mathrm{trained}}$, and are normalized accordingly. Next, we considered a two-dimensional plane spanned by $d_1$ and $d_2$, on which 200 uniformly spaced points are selected with the range from -3 to 3. The loss landscape is constructed by updating the parameters according to
\begin{equation}
    \theta = \theta_{\mathrm{trained}} + \alpha \cdot d_1 + \beta \cdot d_2
    \label{eq: landscapetheta}
\end{equation}
where $\alpha$ and $\beta$ denote the coordinates on the plane. For each point on the plane, the updated parameters are substituted back into the quantum circuit to compute the corresponding loss value. The resulting loss values are plotted along the z-axis, forming a three-dimensional landscape. In particular, the point $(\alpha, \beta) = (0,0)$ corresponds to the trained parameters and the minimum loss value, while the remaining points characterize how the loss changes with respect to parameter update. This visualization provides an intuitive way to analyze the relationship between the cost function and the circuit parameters. 

\section{Result}
\label{sec4}
We evaluate Q-LINK (Fixed), Q-LINK (Adaptive), and the Vanilla model using stochastic gradient descent (SGD) with a learning rate of 0.1. The number of qubits ranges from 5 to 10, including the messenger qubit, and the circuit depth is set to $n^2 \log(n)$ to ensure the circuit is sufficiently complex. For each model, simulation is repeated five times, and the maximum iterations are set to 1500. Convergence is declared when the loss value drops below $10^{-3}$. All models are constructed using TensorCircuit \cite{zhang2023} and PyTorch \cite{paszke2019}, and simulations are performed on a workstation with an AMD Ryzen 9 7960X CPU and an NVIDIA RTX 4090 GPU.

\begin{figure*}[h]
    \centering
    \begin{subfigure}{0.32\textwidth}
        \centering
        \includegraphics[width=\linewidth]{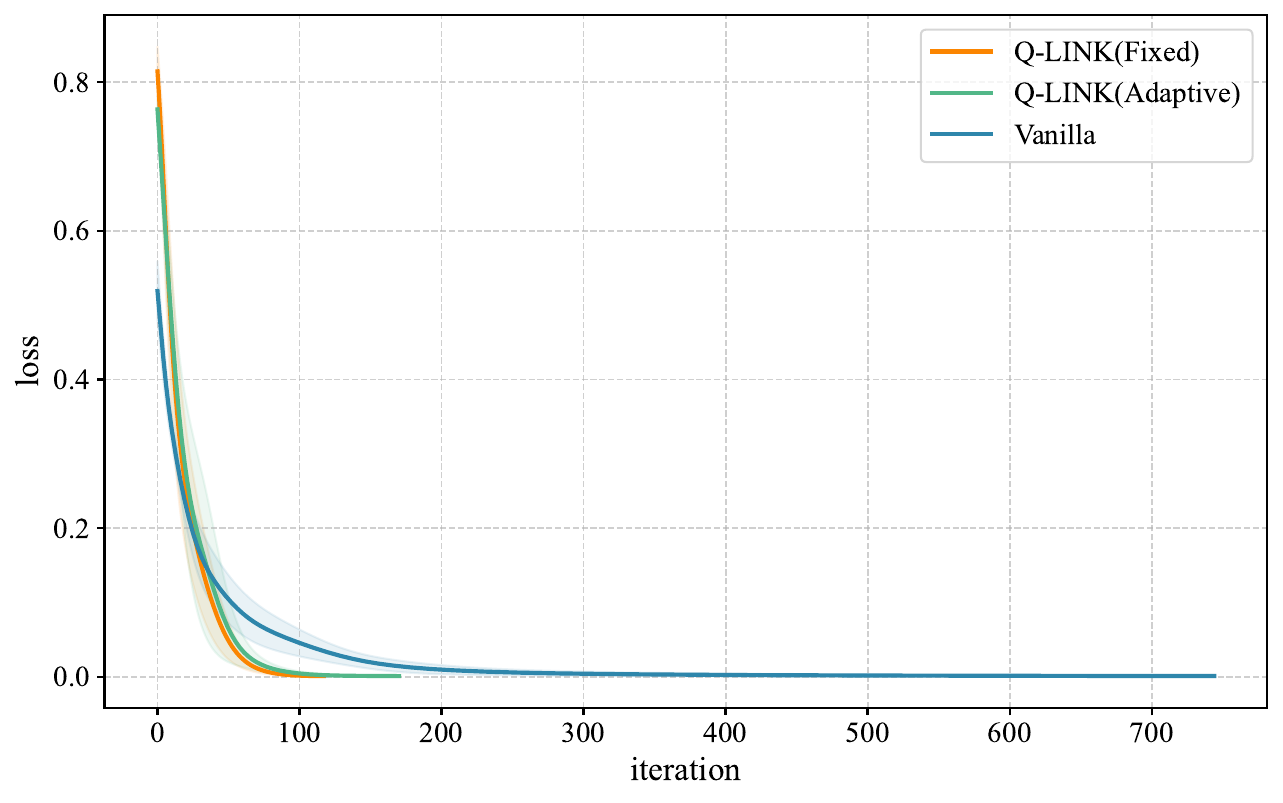}
        \caption{5 Qubits}
    \end{subfigure}
    \hfill
    \begin{subfigure}{0.32\textwidth}
        \centering
        \includegraphics[width=\linewidth]{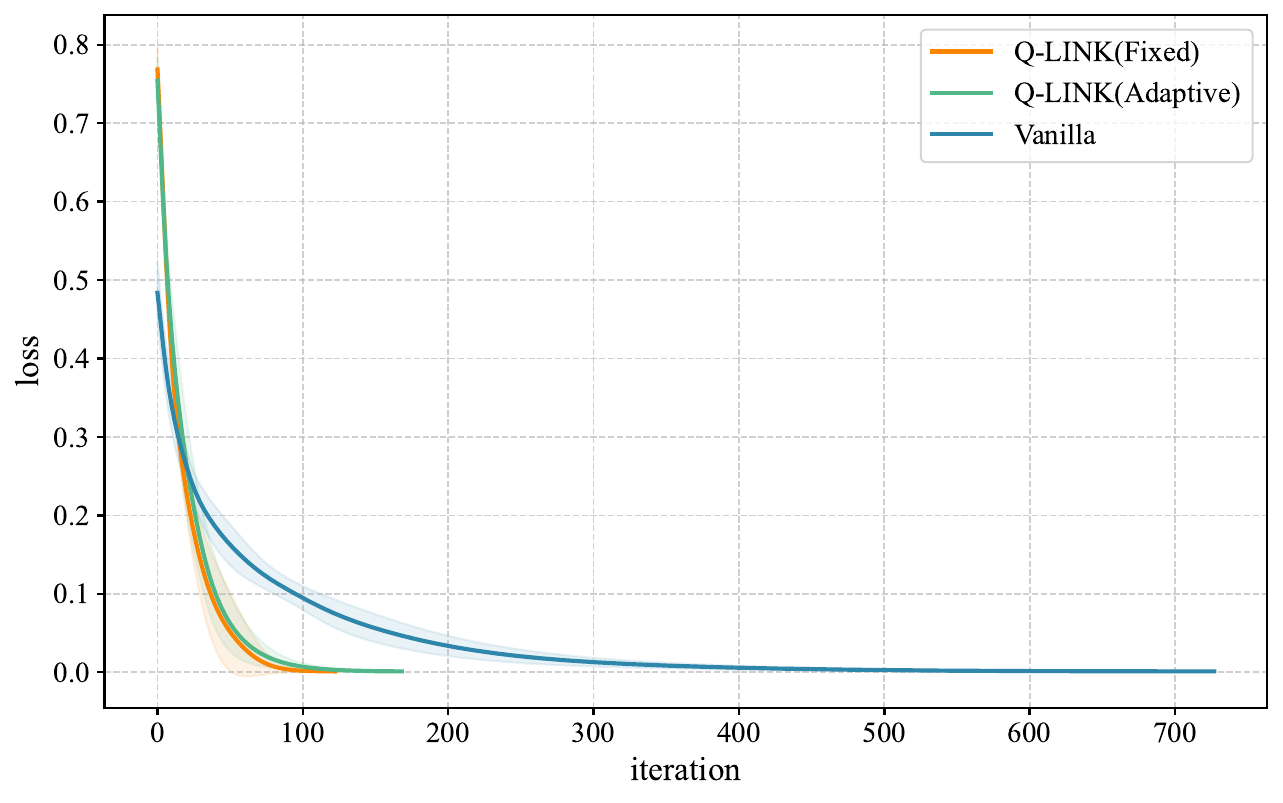}
        \caption{6 Qubits}
    \end{subfigure}
    \hfill
    \begin{subfigure}{0.32\textwidth}
        \centering
        \includegraphics[width=\linewidth]{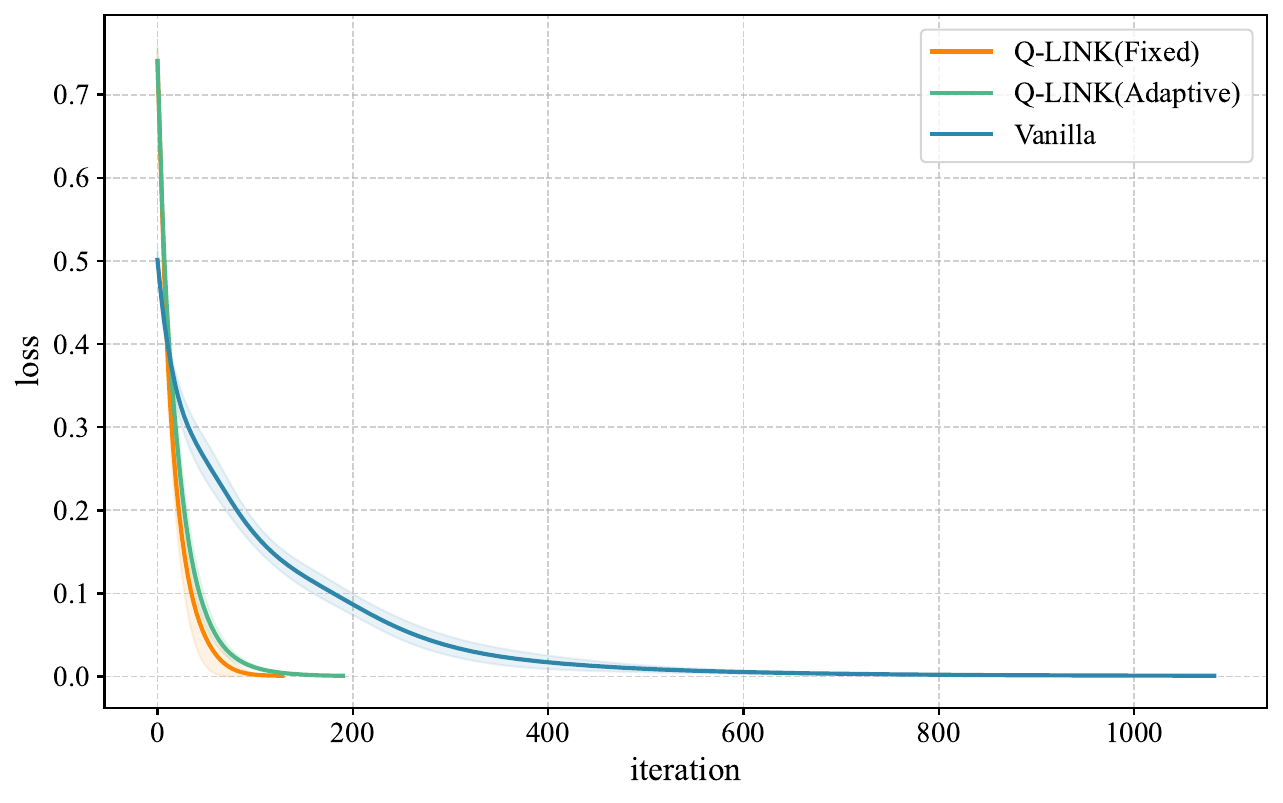}
        \caption{7 Qubits}
    \end{subfigure}

    \vspace{0.3cm}

    \begin{subfigure}{0.32\textwidth}
        \centering
        \includegraphics[width=\linewidth]{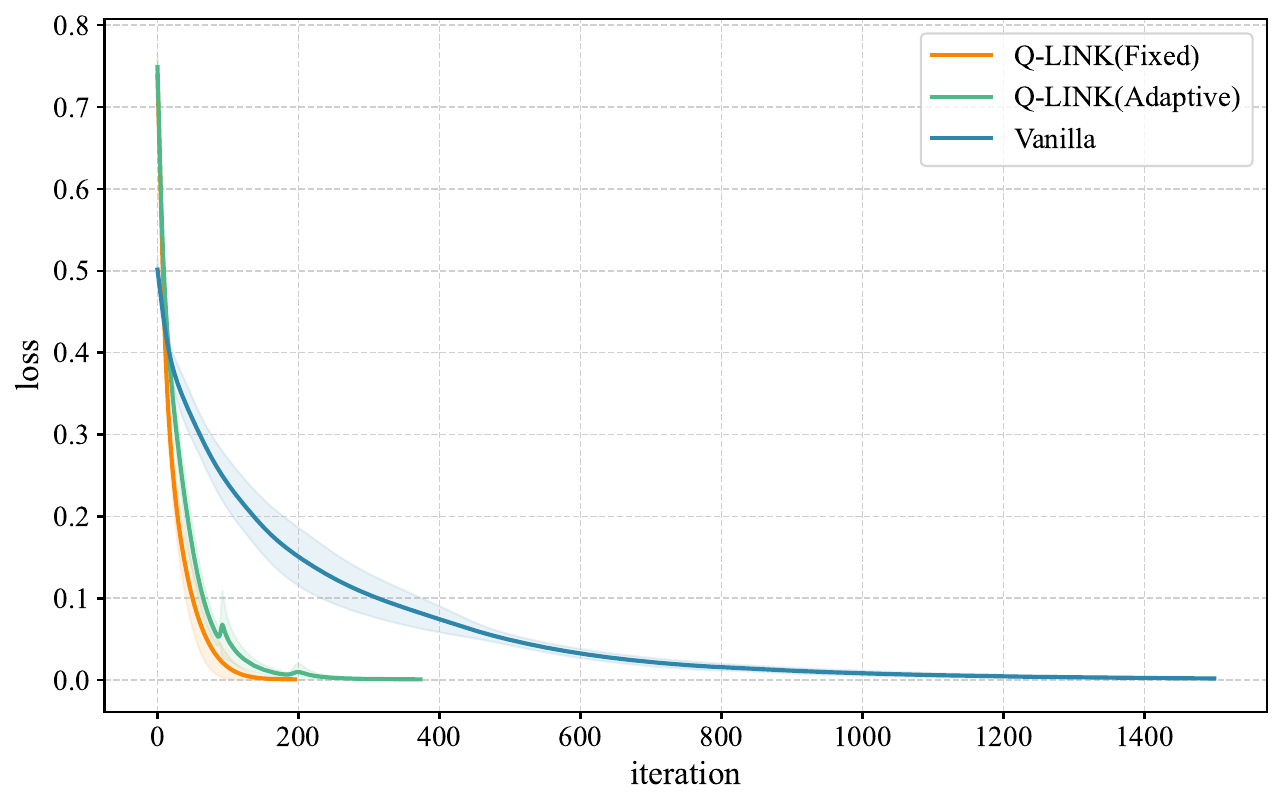}
        \caption{8 Qubits}
    \end{subfigure}
    \hfill
    \begin{subfigure}{0.32\textwidth}
        \centering
        \includegraphics[width=\linewidth]{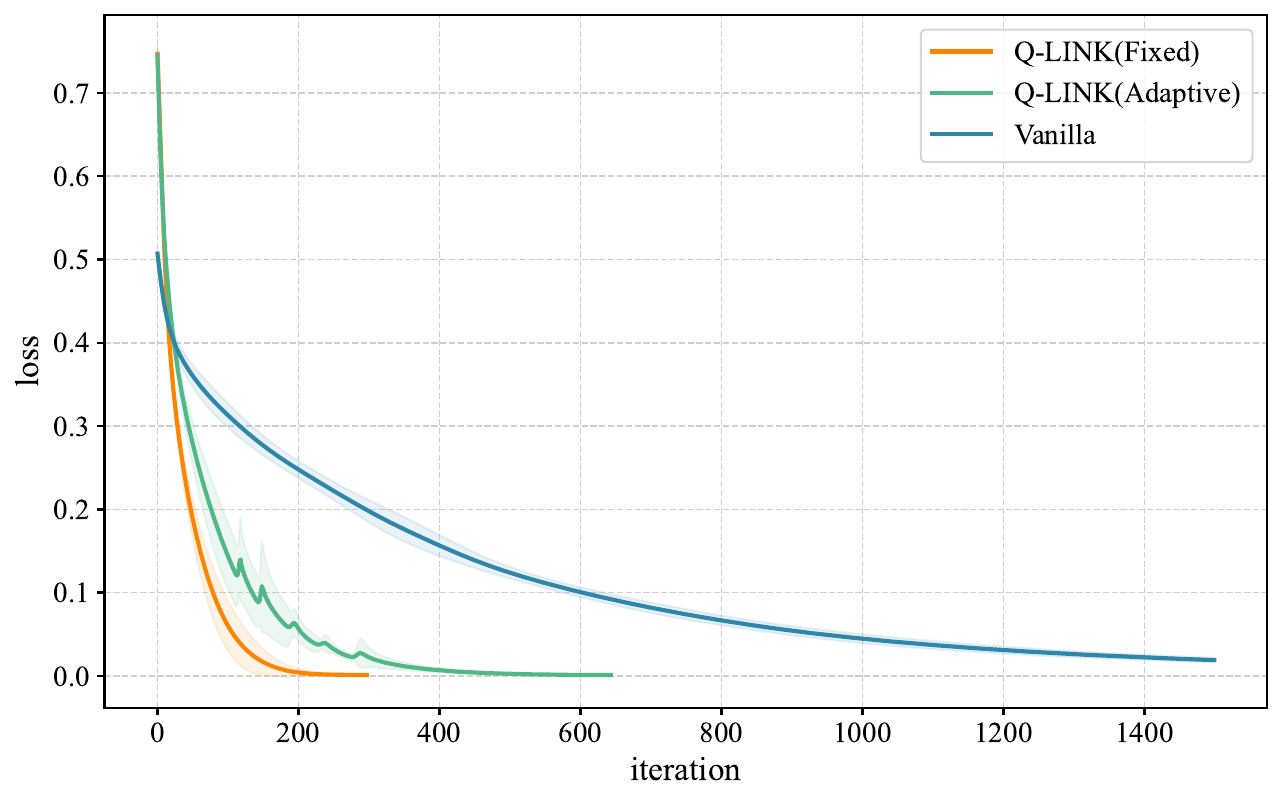}
        \caption{9 Qubits}
    \end{subfigure}
    \hfill
    \begin{subfigure}{0.32\textwidth}
        \centering
        \includegraphics[width=\linewidth]{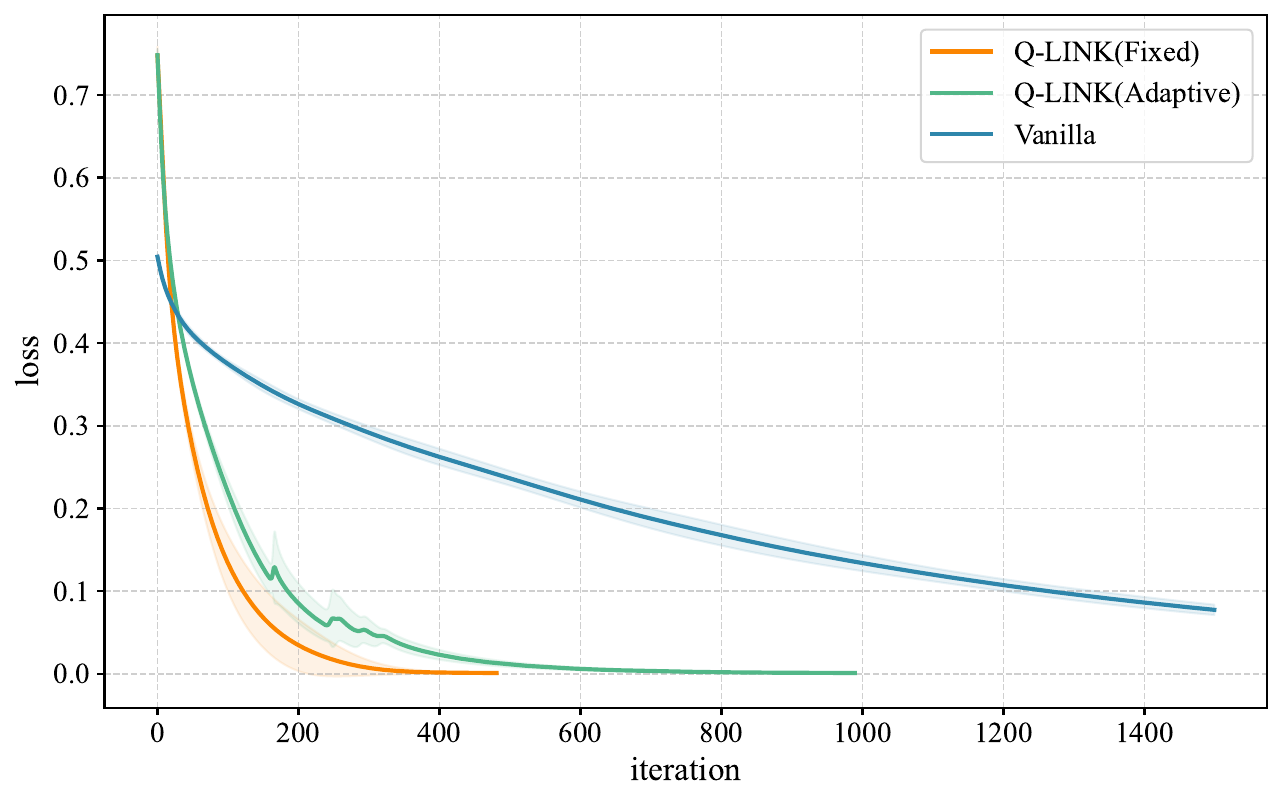}
        \caption{10 Qubits}
    \end{subfigure}

    \caption{Average loss as a function of optimization iteration for different models and different numbers of qubits. The orange, green, and blue curves correspond to Q-LINK (Fixed), Q-LINK (Adaptive), and the Vanilla model, respectively. Solid lines indicate the mean loss over five runs, while the shaded regions represent the standard deviation.}
    \label{fig: lossvsiteration}
\end{figure*}

Figure \ref{fig: lossvsiteration} illustrates the evolution of the loss value with respect to the iterations for different models under different numbers of qubits. Overall, the Q-LINK models converge significantly faster than Vanilla, with Q-LINK (Fixed) exhibiting the best convergence behavior. As the number of qubits increases, Vanilla requires more iterations to converge and fails to converge within the predefined iteration limit starting from eight qubits, which is consistent with barren plateau behavior. Although the convergence speed of Q-LINK also decreases with increasing qubit number, it remains considerably faster than that of Vanilla, suggesting that Q-LINK mitigates the barren plateau effect. Notably, Q-LINK models start from higher initial loss values than Vanilla but still converge more rapidly, indicating more effective exploration of the parameter space. In addition, Q-LINK (Adaptive) shows oscillatory behavior from 8 qubits to 10 qubits, while Q-LINK (Fixed) converges more smoothly.

\begin{figure}
    \centering
    \includegraphics[width=0.7\linewidth]{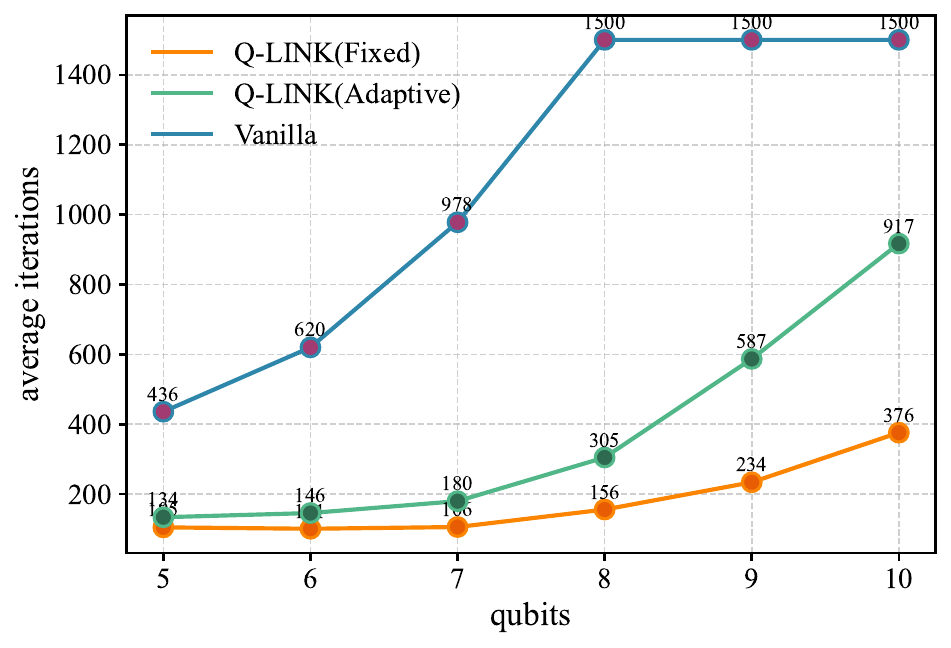}
    \caption{Average number of optimization iterations required for convergence to the target cost function across different numbers of qubits.}
    \label{fig: avgiter_qubits}
\end{figure}

Figure \ref{fig: avgiter_qubits} illustrates the average number of iterations required for convergence under different qubit numbers. We define the ratio between the maximum allowed iterations and the stopping iteration as the convergence efficiency. Q-LINK (Fixed), Q-LINK (Adaptive), and Vanilla achieve convergence efficiencies of 10.55, 6.49, and 1.73, respectively, indicating that Q-LINK converges approximately 10 times more efficiently than Vanilla. As shown in the figure, Q-LINK consistently converges in fewer iterations than Vanilla across all qubit sizes, demonstrating improved convergence efficiency under random quantum state inputs. Among the two Q-LINK variants, the Fixed model requires fewer iterations than the Adaptive model for all qubit numbers. Although the number of required iterations increases with system size, Q-LINK maintains stable convergence behavior, whereas Vanilla exhibits significantly slower convergence. These results suggest that the messenger-qubit residual-like structure effectively mitigates barren plateau effects and enables more scalable optimization.

\begin{figure*}[h]
    \centering
    \begin{subfigure}{0.32\textwidth}
        \centering
        \includegraphics[width=\linewidth]{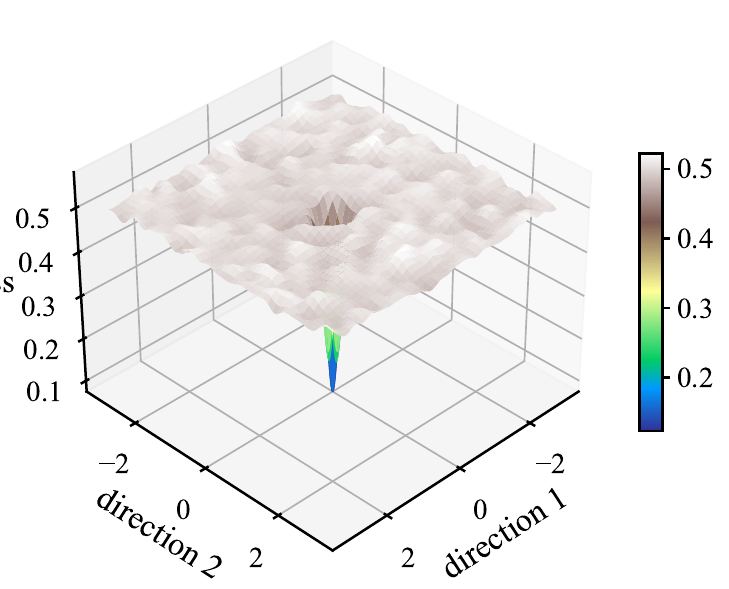}
        \caption{Vanilla (10 qubits)}
    \end{subfigure}
    \hfill
    \begin{subfigure}{0.32\textwidth}
        \centering
        \includegraphics[width=\linewidth]{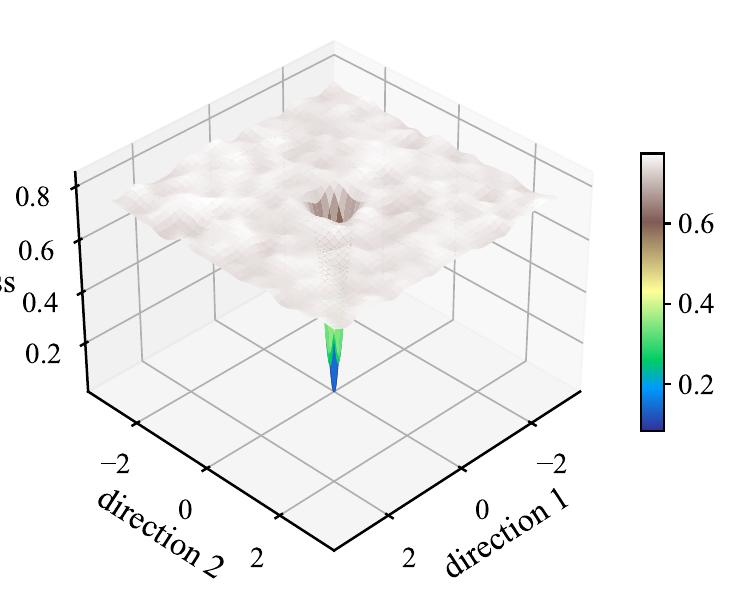}
        \caption{Q-LINK (Fixed) (10 qubits)}
    \end{subfigure}
    \hfill
    \begin{subfigure}{0.32\textwidth}
        \centering
        \includegraphics[width=\linewidth]{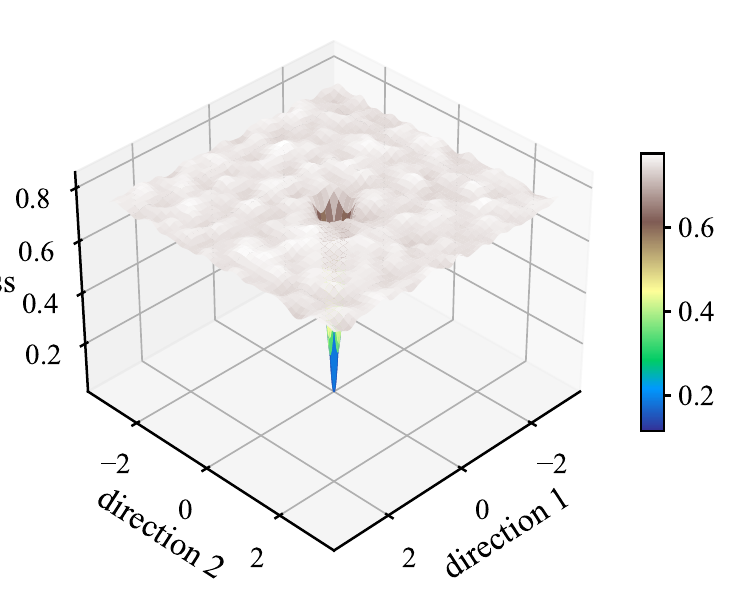}
        \caption{Q-LINK (Adaptive) (10 qubits)}
    \end{subfigure}

    \vspace{0.3cm}

    \begin{subfigure}{0.32\textwidth}
        \centering
        \includegraphics[width=\linewidth]{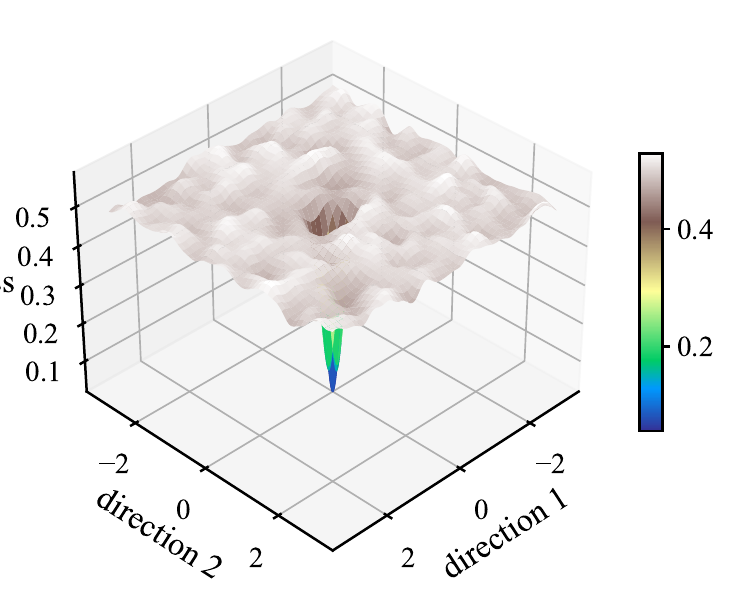}
        \caption{Vanilla (9 qubits)}
    \end{subfigure}
    \hfill
    \begin{subfigure}{0.32\textwidth}
        \centering
        \includegraphics[width=\linewidth]{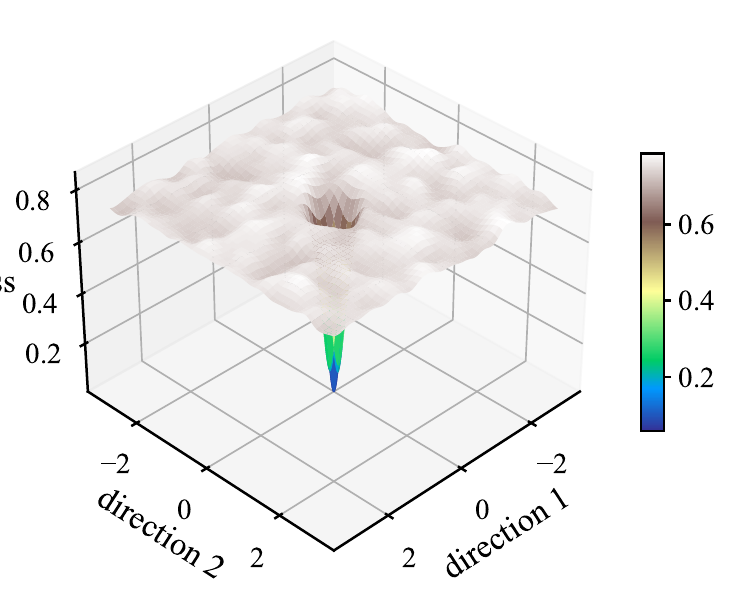}
        \caption{Q-LINK (Fixed) (9 qubits)}
    \end{subfigure}
    \hfill
    \begin{subfigure}{0.32\textwidth}
        \centering
        \includegraphics[width=\linewidth]{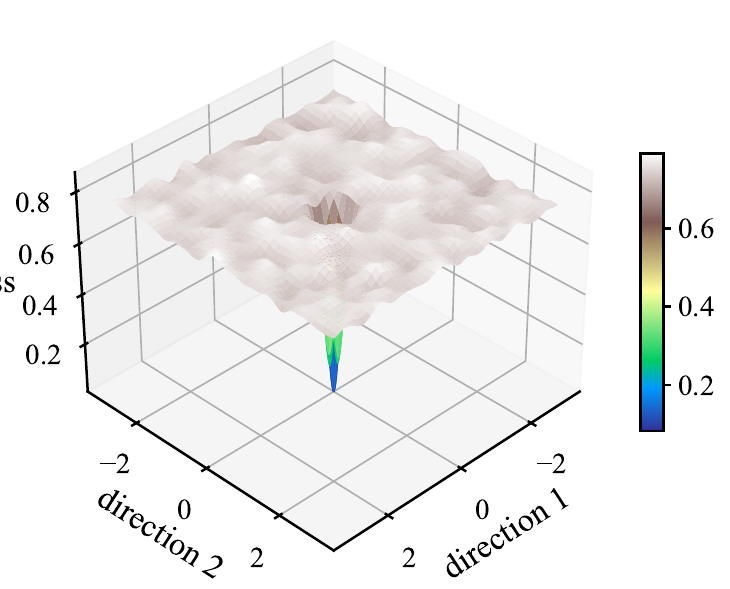}
        \caption{Q-LINK (Adaptive) (9 qubits)}
    \end{subfigure}

     \vspace{0.3cm}

    \begin{subfigure}{0.32\textwidth}
        \centering
        \includegraphics[width=\linewidth]{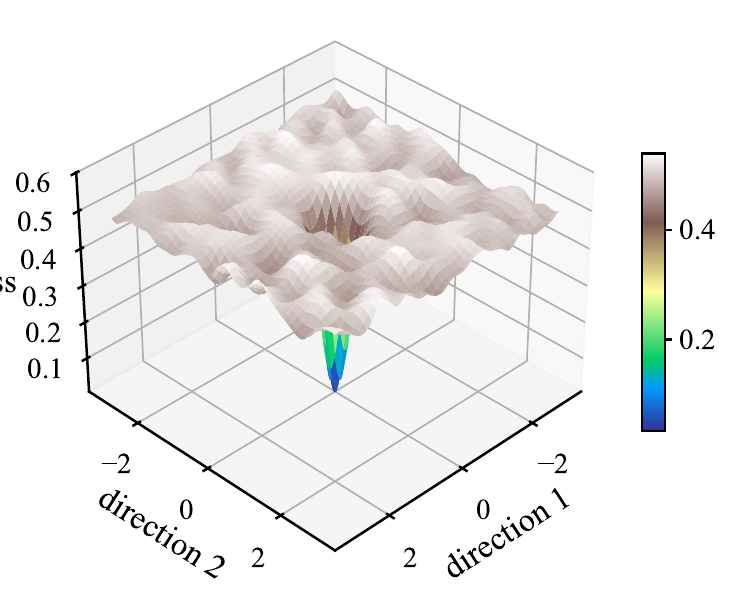}
        \caption{Vanilla (8 qubits)}
    \end{subfigure}
    \hfill
    \begin{subfigure}{0.32\textwidth}
        \centering
        \includegraphics[width=\linewidth]{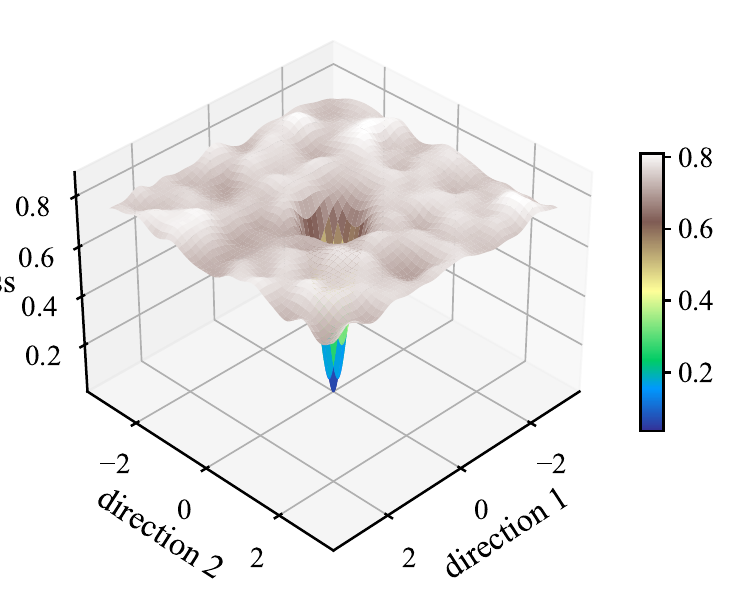}
        \caption{Q-LINK (Fixed) (8 qubits)}
    \end{subfigure}
    \hfill
    \begin{subfigure}{0.32\textwidth}
        \centering
        \includegraphics[width=\linewidth]{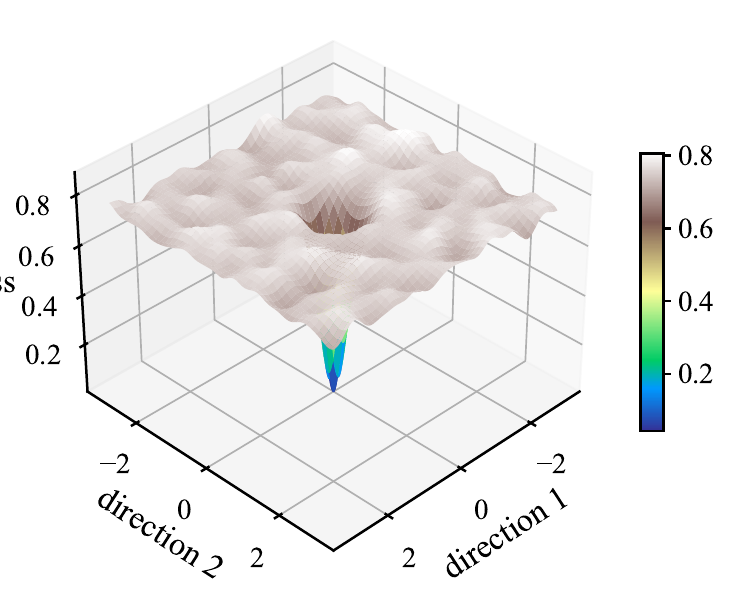}
        \caption{Q-LINK (Adaptive) (8 qubits)}
    \end{subfigure}

    \caption{Loss landscapes of the Vanilla, Q-LINK (Fixed), and Q-LINK (Adaptive) models for 8 to 10 qubits. The landscapes are evaluated along two random parameter directions around the optimal loss value point, where the two horizontal axes denote the parameter directions and the vertical axis denotes the loss value. Subfigures correspond to different models and qubit numbers.}
    \label{fig: losslandscape}
\end{figure*}
Figure \ref{fig: losslandscape} illustrates the loss landscapes of different models, with only the cases for 8, 9, and 10 qubits shown for comparison. As observed from the figure, the loss landscapes of Q-LINK are smoother in the vicinity of the optimal loss value compared to those of the Vanilla model. In contrast, Vanilla exhibits numerous local minima near the optimal loss region, which manifest as hill-like scenario features in the loss landscape. This indicates that the Vanilla model is more likely to become trapped in local minima, thereby requiring more training iterations to achieve the optimal loss. When comparing the two Q-LINK models, the Fixed model exhibits a smoother loss landscape than the Adaptive model, suggesting that Q-LINK (Fixed) can more easily converge to the target loss value. This observation is consistent with the iteration-based convergence results shown earlier. Furthermore, as the number of qubits increases, the number of local minima near the optimal loss region increases for all three models. However, this increase is most pronounced for Vanilla, followed by Q-LINK (Adaptive), while Q-LINK (Fixed) remains comparatively stable. These results further indicate that the Q-LINK (Fixed) architecture can effectively mitigate the barren plateau.

Table \ref{tab: expr_grad} summarizes the expressibility and gradient variance for the Vanilla and the two Q-LINK models. The expressibility values are comparable across all models, indicating similar expressive ability. Therefore, the observed performance differences are not attributed to expressibility, but rather to differences in circuit structure. Regarding gradient variance, Q-LINK (Fixed) exhibits larger variance at small system sizes, where the Q-LINK (Adaptive) dominates at larger qubit numbers. This suggests that adaptive residual helps maintain gradient magnitude as the system scales, although larger variance alone does not guarantee faster convergence, as shown in Figure \ref{fig: avgiter_qubits}. Overall, Q-LINK significantly improves trainability over Vanilla. The values shown as 0 in Table \ref{tab: expr_grad} correspond to the numerical precision, rather than exact zero.
\begin{table}[t]
\centering
\caption{Expressibility and gradient variance for different models across different numbers of qubits.}
\label{tab: expr_grad}
\scalebox{0.7}{
\begin{tabular}{c c c c}
\toprule
\textbf{Qubits} & \textbf{Model} & \textbf{Expressibility} & \textbf{Gradient Variance} \\
\midrule
\multirow{3}{*}{5}
 & Q-LINK(Fixed) & $3.17\times 10^{-3}$ & $6.74\times 10^{-5}$  \\
 & Q-LINK(Adaptive) & $2.26\times 10^{-3}$  & $2.51\times 10^{-5}$  \\
 & Vanilla & $2.89\times 10^{-3}$  & $1.67\times 10^{-5}$  \\

\multirow{3}{*}{6}
 & Q-LINK(Fixed) & $7.55\times 10^{-3}$  & $2.81\times 10^{-5}$  \\
 & Q-LINK(Adaptive) & $3.03\times 10^{-3}$  & $1.25\times 10^{-5}$  \\
 & Vanilla & $3.62\times 10^{-3}$  & $2.74\times 10^{-6}$  \\

\multirow{3}{*}{7}
 & Q-LINK(Fixed) & $1.46\times 10^{-3}$  & $1.51\times 10^{-5}$  \\
 & Q-LINK(Adaptive) & $1.46\times 10^{-3}$  & $4.73\times 10^{-6}$  \\
 & Vanilla & $1.89\times 10^{-3}$  & $1.04\times 10^{-6}$  \\

 \multirow{3}{*}{8}
 & Q-LINK(Fixed) & $2.00\times 10^{-6}$  & $5.38\times 10^{-6}$  \\
 & Q-LINK(Adaptive) & $2\times 10^{-6}$  & $1.60\times 10^{-5}$  \\
 & Vanilla & $8.40\times 10^{-5}$  & $6.02\times 10^{-7}$  \\

 \multirow{3}{*}{9}
 & Q-LINK(Fixed) & $0$ & $2.33\times10^{-6}$ \\
 & Q-LINK(Adaptive) & $0$ & $1.39 \times 10^{-5}$ \\
 & Vanilla & $2\times10^{-6}$ & $2.02 \times 10^{-7}$ \\

 \multirow{3}{*}{10}
 & Q-LINK(Fixed) & $0$ & $9.62 \times 10^{-7}$ \\
 & Q-LINK(Adaptive) & $0$ & $7.27 \times 10^{-6}$ \\
 & Vanilla & $0$ &  $1.18 \times 10^{-7}$ \\
\bottomrule
\end{tabular}
}
\vspace{-5pt}
\end{table}

\section{Conclusion}
\label{sec5}
In this study, we propose the Q-LINK model, a residual-like structure that mitigates the barren plateau in variational quantum algorithms without requiring circuit splitting or intermediate measurements. The proposed approach introduces a single messenger qubit that records information from the data qubits (input states) and redistributes it across circuit layers, enabling effective residual information flow. Using random quantum states as inputs, we evaluate the proposed structure in terms of convergence speed, expressibility, and gradient variance across different numbers of qubits. Both Q-LINK models consistently converge faster than Vanilla. Meanwhile, all models exhibited similar expressibility, indicating that the observed performance improvements were due to the introduced residual structure rather than expressive ability. Furthermore, the fixed and adaptive models exhibit different behavior in gradient variance across different qubit sizes, reflecting a trade-off between optimization efficiency and scalability. These results suggest that introducing an extra single messenger qubit can substantially improve circuit trainability and mitigate the barren plateau. Future work will extend this framework to larger quantum systems and practical application benchmarks, and investigate its performance under noisy quantum environments.

\textbf{Code and Artifacts Availability:} All the artifacts and data used to produce this research are available at {\small\texttt{\kern-0.05em\url{https://github.com/AARC-lab/DCAS_2026_QLINK}}}.

\bibliographystyle{IEEEtran}
\bibliography{reference}

\end{document}